\documentclass[nohyper,paper]{JHEP3}
\usepackage[T1]{fontenc}
\usepackage{epsfig}
\usepackage{graphicx}
\usepackage{axodraw}

\def\i{\'\i}

\def\be{\begin{equation}}
\def\ee{\end{equation}}
\def\bea{\begin{eqnarray}}
\def\eea{\end{eqnarray}}

\def\V{\mbox{V}}

\def\LZ1{\widetilde{{\mathcal Z}}_1}
\def\<{\langle}
\def\>{\rangle}

\def\Z1{\widetilde{Z}_1}

\newcommand{\1}{1\!\!\!\bot}
\newcommand{\hbo}{\hbox to 1 true cm {\hfill } }

\def\gtapprox{\raisebox{0.45ex}{$\,\,>$}
         \raisebox{-0.7ex}{$\!\!\!\!\!\sim\,\,$}}

\title{Numerical Study of the Ghost-Gluon Vertex in Landau gauge}

\author{A.~Cucchieri, T.~Mendes and A.~Mihara \\
        Instituto de F\'{\i}sica de S\~ao Carlos --
        Universidade de S\~ao Paulo \\
        C.P.~369, 13560--970 S\~ao Carlos, SP -- Brazil}

\abstract{We present a numerical study of the ghost-gluon
vertex and of the corresponding renormalization function
$\Z1(p^2)$ in minimal Landau gauge for $SU(2)$ lattice
gauge theory.
Data were obtained for three different lattice volumes
($\V = 4^4, 8^4, 16^4$) and for three lattice couplings 
$\beta = 2.2, 2.3, 2.4$. Gribov-copy effects have
been analyzed using the so-called smeared gauge fixing.
We also consider two different sets of momenta (orbits)
in order to check for possible effects due to the breaking
of rotational symmetry. The vertex has been evaluated at the
asymmetric point $(0;p,-p)$ in momentum-subtraction scheme.
We find that $\Z1(p^2)$ is approximately constant and equal
to 1, at least for momenta $p \gtapprox 1$ GeV. This constitutes
a nonperturbative verification of the so-called nonrenormalization
of the Landau ghost-gluon vertex.
Finally, we use our data to evaluate the running coupling
constant $\alpha_s(p^2)$. }


\preprint{ }

\begin{document}


\section{Introduction}

In the framework of quantum field theory, Faddeev-Popov 
ghosts are introduced in order to quantize non-Abelian gauge
theories. Although the ghosts
are a mathematical artifact and are absent from the physical
spectrum, one can use the ghost-gluon vertex and the ghost
propagator to calculate physical observables, such as the
QCD running coupling $\alpha_s(p^2)$, using the relation
\cite{text}
\be
\alpha_s(p^2) \; = \; \alpha_0\,
  \frac{Z_3(p^2) \, \widetilde{Z}_3^2(p^2)}{\Z1^2(p^2)} \; .
\label{eq:alpha_run}
\ee
Here $\alpha_0 = g_0^2 / 4 \pi $ is the bare coupling constant and
$Z_3(p^2)$, $\widetilde{Z}_3(p^2)$ and $\Z1(p^2)$ are, respectively,
the gluon, ghost and ghost-gluon vertex renormalization functions.
In Landau gauge, the vertex renormalization function $\Z1(p^2)$
is finite and constant, i.e.\ independent of the renormalization scale $ p $,
to all orders of perturbation theory
(see \cite{Taylor:ff} and the nonrenormalization theorems
in \cite[Chapter 6]{Piguet:1995er}). However, a direct nonperturbative
verification of this result is still lacking. This is the purpose
of the present work. Let us stress that, if $\Z1(p^2)$
is finite and constant also at the nonperturbative level, the
equation above can be simplified, yielding
a nonperturbative definition of the running coupling
constant that requires only the calculation of gluon
and ghost propagators
\cite{vonSmekal:1997is,vonSmekal:1997is2}.
A nonperturbative investigation of
the structure of the ghost-gluon vertex is also important
for studies of gluon and ghost propagators using
Dyson-Schwinger equations (DSE) \cite{Alkofer:2000wg}.
In fact, in these studies, one makes use of
Ans\"atze for the behavior of the propagators and vertices
in the equations, in order to obtain solvable truncation schemes.
Of course, a nonperturbative input for these quantities is important for
a truly nonperturbative solution of the DSE.

Let us recall that in Landau gauge the gluon and ghost propagators can be
expressed (in momentum space) as
\bea
D_{\mu \nu}^{b c}(q,-q) &\,=\,& \delta^{b c} \, \left( \delta_{\mu \nu}
        \, - \, \frac{q_{\mu}\,q_{\nu}}{q^2} \right) \, D(q^2)
                         \;=\; \delta^{b c} \, \left( \delta_{\mu \nu}
        \, - \, \frac{q_{\mu}\,q_{\nu}}{q^2} \right) \, \frac{F(q^2)}{q^2}
\label{eq:dprop} \\[2mm]
G^{b c}(q,-q) &\,=\,& - \delta^{b c} \,\, G(q^2)
              \;=\; - \delta^{b c} \,\, \frac{J(q^2)}{q^2}
 \;\mbox{,}
\label{eq:gprop}
\eea
where $F(q^2)$ and $J(q^2)$ are, respectively, the gluon and
ghost form factor and the color indices $ b $ and $ c $ take values
$1, 2, \ldots, N^2_c \,-\,1$ in the $SU(N_c)$ case. Then, in 
the momentum-subtraction scheme one has that the gluon and ghost
renormalization functions are given by
\bea
F_R(q^2, p^2)  &\,=\,& Z_{3}^{-1}(p^2) \, F(q^2)
                                             \label{eq:Drenor} \\[2mm]
J_R(q^2, p^2)  &\,=\,& \widetilde{Z}_{3}^{-1}(p^2) \, J(q^2) 
\label{eq:Grenor}
\eea
with the renormalization conditions
\be
F_R(p^2, p^2) \; = \; J_R(p^2, p^2) \; = \; 1 \; .
\ee
Thus, if one can set $\Z1(p^2) \,=\, 1$, the
running coupling (\ref{eq:alpha_run}) can be written as
\cite{vonSmekal:1997is,vonSmekal:1997is2}
\be
\alpha_s(p^2) \; = \; \alpha_0\, F(p^2)\,J^2(p^2) \; ,
\label{eq:alpha_run2}
\ee
i.e.\ one only needs to evaluate the gluon and ghost form factors
defined above.
In recent years, the infrared behavior of these form factors
has been extensively studied (in Landau gauge) using different
analytical approaches
\cite{vonSmekal:1997is,vonSmekal:1997is2,Atkinson:1997tu,Atkinson:1998zc,
Zwanziger:2001kw,Bloch:2001wz,Lerche:2002ep,Fischer:2002hn,
Fischer:2002eq,Alkofer:2003jj,Alkofer:2003jk,Zwanziger:2002ia,Bloch:2003yu,
Pawlowski:2003hq,Kondo:2003sw,Kondo:2003rj,Sobreiro:2004us,
Sobreiro:2004yj,Aguilar:2004thesis,Aguilar:2004kt,Fischer:2004uk}.
Also, numerical studies of these form factors and of the
running coupling defined in eq.\ (\ref{eq:alpha_run2}) have been reported in
\cite{Cucchieri:2001za,Bloch:2002we,Langfeld:2002dd,Bloch:2003sk} for the
$SU(2)$ group and in
\cite{Leinweber:1998uu,Bonnet:2001uh,
Nakajima:2002kh,Nakajima:2003sg,Furui:2003jr,
Nakajima:2003my,Furui:2003mz,Furui:2004bq} for the $SU(3)$ case.

An indirect evaluation of the ghost-gluon vertex renormalization function 
has been recently presented in \cite{Bloch:2003sk}, confirming
that $\Z1(p^2)$ is finite in the continuum limit.
On the other hand, a direct nonperturbative verification of this
result from a numerical evaluation of the ghost-gluon vertex
is still missing. Let us stress that a direct evaluation of $\Z1(p^2)$ 
would allow a study of the running coupling constant
\cite{Furui:2000mq} using
eq.\ (\ref{eq:alpha_run}) instead of eq.\ (\ref{eq:alpha_run2}).
Such a study may improve the precision of the determination of
$\alpha_s(p^2)$, since in that case one does not need, in principle, to
use the so-called {\em matching rescaling} technique \cite{Bloch:2003sk}
when considering data obtained at different $\beta$ values.

In this paper we study numerically the
ghost-gluon vertex and the corresponding renormalization
function $\Z1(p^2)$ for the $SU(2)$ case in the minimal
Landau gauge. The definition of this renormalization function
(in the continuum) is presented in
Section \ref{sec:vertex}.
Numerical simulations are explained
in Section \ref{sec:data}. In particular, in Section
\ref{sec:vertexlattice}
we define the vertex renormalization function on the
lattice and compare our direct determination of $\Z1(p^2)$ to
the indirect evaluation presented in \cite{Bloch:2003sk}.
Note that, since the numerical evaluation of the ghost-gluon vertex
may be affected by Gribov-copy effects (see discussion in
Sections \ref{sec:data} and \ref{sec:results} below), we evalute this quantity
using two different gauge-fixing methods, leading to two different
sets of Gribov copies.
A comparison of the results obtained in the two cases
allows us to estimate the influence of Gribov
copies (the so-called Gribov noise) on the considered
quantity.

Results for the ghost-gluon vertex
renormalization function, the gluon and ghost
propagators and the running coupling
constant $\alpha_s(p^2)$ are reported in Section \ref{sec:results}.
Finally, in Section \ref{sec:conclusions}
we draw our conclusions. Preliminary results have been reported
in \cite{Mihara}.


\section{The ghost-gluon vertex}
\label{sec:vertex}

\FIGURE[t]{
\parbox{5.8in}{
\hskip 0.9in
\epsfig{file=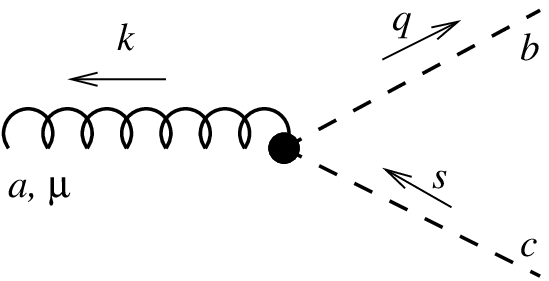,width=10cm}}
\caption{Notation (momenta and indices) for the ghost-gluon vertex.
\label{fig:vert}}}

Following the notation in Ref.\ \cite{Alkofer:2000wg} (see also
Fig.\ \ref{fig:vert}) the 3-point function for
$A_{\mu}^a$ (gluon), $\eta^b$ (ghost) and
$\bar{\eta}^c$ (anti-ghost) fields
--- corresponding to the ghost-gluon vertex ---
is given by
\be 
V^{abc}_{\mu}(x,y,z) \; = \; \langle \, A^{a}_{\mu}(x) \, \eta^{b}(y) 
\, \bar{\eta}^{c}(z) \, \rangle \, \, .
\ee
Going to momentum space and using translational invariance for the
3-point function, one gets
\bea
V^{abc}_{\mu}(k;q,s) & = &
\int d^4x\, d^4y\, d^4z\; e^{-i(kx + qy - sz)} \; V^{abc}_{\mu}(x,y,z) \\[2mm]
&=& \int d^4z\; e^{-i(k+q-s)z} \int d^4x\, d^4y\; e^{-i(kx + qy)}
\;\langle \, A^{a}_{\mu}(x)
  \, \eta^{b}(y) \, \bar{\eta}^{c}(0) \, \rangle \label{eq:eeA}  \\[2mm]
 & = & (2 \pi)^4 \; \delta^4(k+q-s) \;
                              G^{abc}_{\!\mu}(k,q) \,\, ,
\label{eq:V}
\eea
where $\delta^4(k+q-s)$ implies conservation of momentum. 
Then, the ghost-gluon vertex function is obtained
by ``amputating'' the corresponding 3-point function (see Fig.\ \ref{fig:3pt})
\be
\Gamma^{abc}_{\!\mu}(k,q) \; = \;
\frac{G^{abc}_{\!\mu}(k,q)}{D(k^2) \, G(q^2) \, G(s^2)} \,\, ,
\label{eq:lamb}
\ee
where $s = k+q$ and the functions
$D(k^2)$ and $G(q^2)$ have been defined in eqs.\
(\ref{eq:dprop}) and (\ref{eq:gprop}), respectively.
At tree level (in the continuum) one obtains \cite{text}
\be
\Gamma^{abc}_{\!\mu}(k,q) \; = \; i \, g_0 \, f^{abc} \, q_{\mu} \,\, ,
\label{eq:gamma0lev}
\ee
where $g_0$ is the bare coupling and $f^{abc}$ are the structure functions
of the $SU(N_c)$ Lie algebra.
This implies the well-known result that the ghost-gluon vertex
is proportional to the momentum $\,q\,$ of the outgoing ghost.
More generally, we can write the relation \cite{Alkofer:2000wg}
\be
\Gamma^{abc}_{\!\mu}(k,q) \; = \; g_0\, f^{abc} \, \Gamma_{\mu}(k,q) \,\, ,
\label{eq:G}
\label{eq:g_gamma}
\ee
where $\Gamma_{\mu}(k,q)$ is the so-called reduced vertex function.
Then, multiplying the previous equation by $f^{dbc} \, \delta^{ad}$,
summing over $a, b, c, d$ and using the relation $ \sum_{b, c} \,
f^{dbc} f^{abc} = N_c \, \delta^{da}$ we get
\be
\Gamma_{\mu}(k,q) \; = \; \frac{1}{g_0 \, N_c  \,(N_c^2 \, - \, 1)} \,
              \sum_{a,b,c}\, f^{abc}\,\Gamma^{abc}_{\!\mu}(k,q) \,\, .
\ee
Since at tree level
one has $\Gamma_{\mu}(k,q) = i \, q_{\mu}$ we can also write
\be
\Gamma_{\mu}(k,q) \; = \; i \, q_{\mu} \, \Gamma(k^2,q^2)
\ee
and it is the scalar function $\Gamma(k^2,q^2)$ that gets renormalized,
when considering a given renormalization scheme.

\FIGURE[t]{
\parbox{5.8in}{
\hskip 0.8in
\epsfig{file=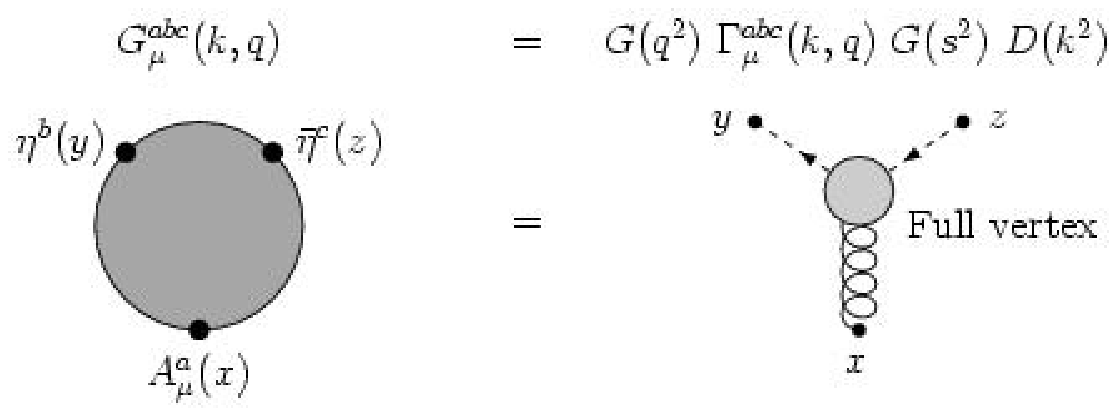,width=11cm}}
\caption{The 3-point function $G_{\mu}^{abc}(k,q)$ and its
relation with the full ghost-gluon vertex $\Gamma_{\mu}^{abc}(k,q)$.
\label{fig:3pt}}}

Clearly, from the above relations we obtain
\be
\Gamma(k^2,q^2) \; =\; \frac{-i}{q^2} \, \sum_{\mu}\, q_{\mu}\,
                         \Gamma_{\mu}(k,q) \; = \; 
             \frac{-i}{g_0 \, N_c  \,(N_c^2 \, - \, 1)}\,
                      \frac{1}{q^2}\,
              \sum_{\mu}\,q_{\mu}\,
              \sum_{a,b,c}\, f^{abc}\,\Gamma^{abc}_{\!\mu}(k,q) \,\, .
\label{eq:gammak2q2}
\ee

In our simulations we considered the asymmetric point
with zero momentum for the gluon ($k = 0$), implying $s = q$.
Formula (\ref{eq:gammak2q2}) then becomes
\be
\Xi(q^2) \;=\; \Gamma(0,q^2) \; =\; 
           \frac{-i}{g_0 \, N_c  \,(N_c^2 \, - \, 1)}\,\frac{1}{q^2}\,
              \sum_{\mu}\,q_{\mu}\,
              \sum_{a,b,c}\, f^{abc}\,\Gamma^{abc}_{\!\mu}(0,q) \,\, .
\label{eq:xi}
\ee
Let us note that the factor $- i$ in the equations above
disappears, since only the
imaginary part of the vertex function $\,\Gamma^{abc}_{\!\mu}(0,q) \!$
contributes to the quantity $\,\Xi(q^2)\,$, i.e.\ we can write
\be
\Xi(q^2) \;=\; \frac{1}{g_0 \, N_c  \,(N_c^2 \, - \, 1)} \;
     \frac{1}{q^2}\; \sum_{\mu}\; q_{\mu} \,
      \sum_{a,b,c}\, f^{abc}\;
      \mbox{Im}\; \Gamma^{abc}_{\!\mu}(0,q) \,\, .
\ee

Finally, in momentum-subtraction scheme one fixes the vertex renormalization
function $ \Z1(p^2) $ by requiring (see also Fig.\ \ref{fig:Z1}) 
\bea
{\Xi_{R}(q^2,p^2)} &=& \Z1(p^2) \, \Xi(q^2) \\[2mm]
{\Xi_{R}(p^2,p^2)} &=& 1 \,\, ,
\label{eq:Z1ren}
\eea
namely the renormalized reduced vertex function
$\Gamma_{\mu}(0,q)$ is equal to the tree-level value
$\,i\,q_{\mu}\,$ at the renormalization scale.
Then, the vertex renormalization function is given by
\be
 \Z1^{-1}(p^2) \; = \; \Xi(p^2) \,\, .
\label{eq:z1L}
\ee
We also obtain that the running coupling can be written as
\bea
g_s(p^2) &=& g_0 \; \Z1^{-1}(p^2) \, Z_3^{1/2}(p^2) \,
                \widetilde{Z}_3(p^2) \\[2mm]
       &=& g_0 \; \Xi(p^2) \, F^{1/2}(p^2) \, J(p^2) \\[2mm]
       &=& \frac{1}{N_c \, (N_c^2 \, - \, 1)} \,
            \frac{F^{1/2}(p^2) \, J(p^2)}{p^2\,D(0) \, G^2(p^2)} \,
                \sum_{\mu}\,p_{\mu} \,\, \sum_{a,b,c}\,
         f^{abc}\; \mbox{Im}\; G_{\mu}^{abc}(0,p) \\[2mm]
  &=& \frac{1}{N_c \, (N_c^2 \, - \, 1)} \,
            \frac{F^{1/2}(p^2)}{D(0) \, G(p^2)} \,
                \sum_{\mu}\,p_{\mu} \,\, \sum_{a,b,c}\,
         f^{abc}\; \mbox{Im}\; G_{\mu}^{abc}(0,p)
            \,\, ,
\eea
where we used eq.\ (\ref{eq:gprop}).

\FIGURE[t]{
\parbox{5.8in}{
\hskip 0.75in
\epsfig{file=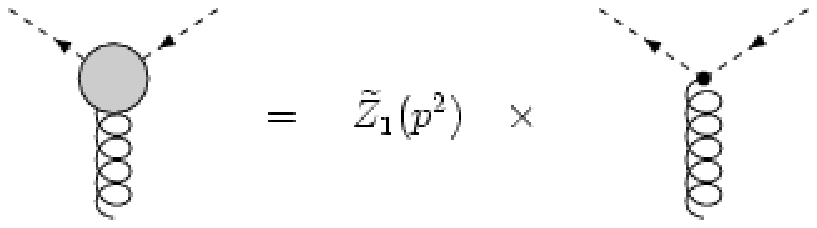,width=11cm}}
\caption{Relation between the full (left) and the bare (right) vertices.
\label{fig:Z1}}}


\section{Numerical Simulations}
\label{sec:data}

For the simulations (see \cite{Cucchieri:2003zx} for details)
we consider the standard Wilson action,
thermalized using heat-bath \cite{Creutz:zw} accelerated by
{\it hybrid overrelaxation} \cite{Adler:1988gc,Brown:1987rr,Wolff:1992ri}.
Since we are considering the $SU(2)$ gauge group we
parametrize the link variables $ U_\mu(x) $ as
\be
U_\mu(x) \; = \; u_\mu^0(x) \,\1 \, + \, i \, \vec{u}_\mu(x) \,
\vec{\sigma} \,\, ,
\ee
where $\1$ is the $2 \times 2$ identity matrix,
$\sigma^{b}$ are the Pauli matrices and the relation
\be
\bigl[ u^0_\mu(x) \bigr]^2 \, + \,
\bigl[ \vec{u}_\mu(x) \bigr]^2 \;=\; 1
\ee
is satisfied. We also consider the standard
definition of the lattice gluon field, i.e.\
\be
A_{\mu}(x) \;=\; A_{\mu}^{b}(x)\,\sigma^{b}\;=\;
  \frac{U_\mu(x) \,-\, U_{\mu}^{\dagger}(x)}{2 \, i} \; ,
\label{eq:gluon}
\ee
implying that the lattice gluon field component $A^b_{\mu}(x)$ is
equal to $ u^b_{\mu}(x)$.
The connection between the lattice link variables $ U_\mu(x) $ and
the continuum gauge field $A_{\mu}(x)$ is given by
\be
U_{\mu}(x) \; = \; \exp{\left[ i \, a\, g_0 \, 
                        A^b_{\mu}(x)\, t^b \right]} \; ,
\label{eq:UandA}
\ee
where $ t^b = \sigma^b /2$ are the generators of the SU(2) algebra,
$\,a\,$ is the lattice spacing and $\,g_0$ is the
bare coupling constant.
Then, the lattice quantity $\,2\,A^b_{\mu}(x) / ( a \, g_0 )\,$
approaches the continuum-gluon-field component
$ A^b_{\mu}(x) $ in the naive continuum limit $\,a \to 0\,$.
Let us also recall that a generic lattice
momentum $\hat{q}$ has components (in lattice units)
\be
\hat{q}_{\mu} = 2 \,
   \sin{\left(\frac{\pi\, \widetilde{q}_{\mu} \, a}{L_{\mu}}\right)} \,\, .
\label{eq:platt}
\ee
Here $L_{\mu} = a \, N_{\mu}$ is the
physical size of the lattice in the $\mu$ direction,
the quantity $\widetilde{q}_{\mu}$
takes values $\lfloor \, \frac{- N_{\mu}}{2} \,\rfloor +
1\, , \ldots , \lfloor \, \frac{N_{\mu}}{2} \, \rfloor \,$ and
$N_{\mu}$ is the number
of lattice points in the $\mu$ direction. If we keep
the physical size $L_{\mu} $ constant and
indicate with $q_{\mu}$ the momentum components in the continuum, we
find that the lattice components
$ \hat{q}_{\mu} = a \, q_{\mu} $ take values in the
interval $( -\pi , \, \pi ]$ when $ a \to 0 $.

\TABULAR[t]{|c|c|c|c|}{
\hline 
$\,\,$ Lattice Volume $V = N^4$ $\,\,$ &
             \( 4^{4} \) & \( 8^{4} \)& \( 16^{4} \)   \\
\hline
$\,\,$ No.\ of Configurations $\,\,$ & $\,\,$ 1000 $\,\,$ & $\,\,$
                       400 $\,\,$ & $\,\,$ 250 $\,\,$ \\
\hline
}{Lattice volumes and total number of configurations
considered. \label{table:data}}

In Table \ref{table:data} we show the lattice volumes used in
the simulations and the corresponding numbers of
configurations. For each volume we
have considered three values of the lattice
coupling, namely $\beta = 2.2, 2.3, 2.4$. The corresponding
string tensions in lattice units \cite{Bloch:2003sk} are, respectively,
equal to $\,\sigma\, a^2 =  0.220(9), 0.136(2)$ and $0.071(1)$.

The minimal (lattice) Landau gauge is implemented using the
stochastic overrelaxation algorithm
\cite{Cucchieri:1995pn,Cucchieri:1996jm,Cucchieri:2003fb}.
We stop the gauge
fixing when the average value of $ \, (\nabla \cdot A)^{2} \,$
--- see Eq.\ (6.1) in \cite{Cucchieri:2003fb} for a definition ---
is smaller than $10^{- 13}$.
The gluon and ghost propagators in momentum space
are evaluated using the relations
\bea
D(0) \,&=&\, \frac{1}{12}\,\sum_{b, \mu}\,
                        D_{\mu \mu}^{\,\!\,b b}(0) \\
D(\widetilde{q}) \,&=&\, \frac{1}{9}\,\sum_{b, \mu}\,
                        D_{\mu \mu}^{\,\!\,b b}(\widetilde{q}) \\
G(\widetilde{q}) \,& =&\,
          \frac{1}{3} \,\sum_{b} \, G^{\,\!\,b b}(\widetilde{q}) \; ,
\eea
where
\bea
D_{\mu \nu}^{b c}(\widetilde{q}) \,& = &\, \frac{1}{V} \, \sum_{x, y}
\, \exp{\left[ \, i\, \widetilde{q} \left(x - y\right) \,\right]} \,
\bigl\langle \, A^b_\mu(x) \, A^c_\nu(y) \, \bigr\rangle \; , \\[2mm]
G^{b c}(\widetilde{q}) \,& = &\, \langle 
\,\left(M^{-1}\right)^{bc}(\widetilde{q}) \, \rangle \; ,
\label{eq:Gofp}  \\[2mm]
\left(M^{-1}\right)^{bc}(\widetilde{q}) \,& = &\, \frac{1}{V} \,
        \sum_{x, y} \; \exp \left[\,- i\, \widetilde{q}
                      \left(x - y\right) \, \right] \;
                \left(M^{-1}\right)^{bc}_{xy} \; ,
        \label{eq:mm1}
\eea 
$V$ is the lattice volume and $ M^{bc}_{xy} $ is a lattice
discretization of the Faddeev-Popov operator $\left[ (-\partial + A) \cdot
\partial \right] \,$. The expression of
this matrix in terms of the gauged-fixed link variables can be
found in \cite[eq.\ (B.18)]{Zwanziger:1993dh} for the generic
$SU(N_c)$ case and in
\cite[eq.\ (11)]{Cucchieri:1997ns} for the $SU(2)$ case considered
in the present work. Let us recall that
this matrix has a trivial null eigenvalue corresponding to a
constant eigenvector. Thus, one can evaluate the inverse of
$ \,M^{bc}_{xy} \, $ only in the space orthogonal to constant vectors,
i.e.\ at non-zero momentum. For the inversion we used a
conjugate gradient method with even/odd preconditioning.

\TABULAR[t]{|c|c|c|c|}{
\hline
      & $\,\,\,\,$ V = \( 4^{4} \) $\,\,\,\,$ 
      & $\,\,\,\,$ V = \( 8^{4} \) $\,\,\,\,$ 
      & $\,\,\,\,$ V = \( 16^{4} \) $\,\,\,\,$ \\
\hline
$\,\,\beta =$ 2.2 $\,\,$ & $\,\,$ 4.3 $\%$ $\,\,$ & $\,\,$ 25.0 $\%$ $\,\,$ &
                    $\,\,$ 97.2 $\%$ $\,\,$ \\   
$\,\,\beta =$ 2.3 $\,\,$ & $\,\,$ 4.4 $\%$ $\,\,$ & $\,\,$ 16.8 $\%$ $\,\,$ &
                    $\,\,$ 72.8 $\%$ $\,\,$ \\   
$\,\,\beta =$ 2.4 $\,\,$ & $\,\,$ 4.1 $\%$ $\,\,$ & $\,\,$ 9.8 $\%$ $\,\,$ &
                    $\,\,$ 50.8 $\%$ $\,\,$ \\   
\hline
}{Percentage of configurations for which the smeared gauge fixing
has found a (different) Gribov copy. Results are reported for the three
lattice volumes and the three $\beta$ values considered.
\label{table:gribov}}

\TABULAR[b]{|c|c|c|c|}{
\hline
      & $\,\,\,\,$ V = \( 4^{4} \) $\,\,\,\,$ 
      & $\,\,\,\,$ V = \( 8^{4} \) $\,\,\,\,$ 
      & $\,\,\,\,$ V = \( 16^{4} \) $\,\,\,\,$ \\
\hline
$\,\,\beta =$ 2.2 $\,\,$ & $\,\,$ 46.5 $\%$ $\,\,$ & $\,\,$ 52.0 $\%$ $\,\,$ &
                    $\,\,$ 63.8 $\%$ $\,\,$ \\   
$\,\,\beta =$ 2.3 $\,\,$ & $\,\,$ 38.6 $\%$ $\,\,$ & $\,\,$ 52.2 $\%$ $\,\,$ &
                    $\,\,$ 65.4 $\%$ $\,\,$ \\   
$\,\,\beta =$ 2.4 $\,\,$ & $\,\,$ 39.0 $\%$ $\,\,$ & $\,\,$ 48.7 $\%$ $\,\,$ &
                    $\,\,$ 65.4 $\%$ $\,\,$ \\   
\hline
}{Percentage of (different)
Gribov copies for which the smeared gauge fixing
has found a smaller value of the minimizing function.
Results are reported for the three
lattice volumes and the three $\beta$ values considered.
\label{table:gribov2}}

Finally, in order to check for possible Gribov-copy effects
we have done (for each thermalized configuration)
a second gauge fixing using the smearing method 
introduced in \cite{Hetrick:1997yy}.
To this end we have applied the APE smearing process
\be
U_{\mu}(x) \,\to\,  (1\,-\,w)\, U_{\mu}(x) \,+\, w \, \Sigma_{\mu}^{\dagger}(x)
\; ,
\ee
where $\,\Sigma_{\mu}(x)\,$ represents the sum over the connecting staples
for the link $\,U_{\mu}(x)$,
followed by a reunitarization of the link matrix. We have chosen
$w = 10/6$ and stopped the smearing when the condition
\be
\frac{Tr}{2} \, \overline{W}_{1,1} \, \geq \, 0.995
\ee
was satisfied. (This is usually achieved with a few
APE-smearing sweeps.) Here $ W_{1,1} $ is the $1 \times 1$ loop and
the average is taken over all $ W_{1,1} $ loops of a given
configuration.
The gauge fixing for the smeared configuration as well as the
final gauge-fixing step have been done again using the stochastic
overrelaxation algorithm. The smeared gauge-fixing method is supposed to
find a unique Gribov copy, even though
this copy does not always correspond to the absolute minimum
of the minimizing functional \cite{Hetrick:1997yy}.
In our simulations we found that the number of
{\em different} Gribov copies obtained with the smeared gauge fixing
increases with larger lattice volumes and with smaller
$\beta$ values (see Table \ref{table:gribov}).
This is in agreement with previous studies
\cite{Marenzoni:1993gh,Cucchieri:1997dx,Cucchieri:1997ug} and
it should be related to an increasing number of local minima for the
minimizing function when the system is highly disordered.
We also found that the minimum obtained with the smeared
gauge fixing is not always smaller than the one obtained without
smearing, but at larger lattice volumes the smearing approach
seems to be more effective in getting closer to the
absolute minimum of the minimizing function (see Table \ref{table:gribov2}).

\subsection{Ghost-Gluon Vertex on the Lattice}
\label{sec:vertexlattice}

On the lattice, the definition of the vertex renormalization function
is obtained as was done in the continuum (see Section \ref{sec:vertex}).
The only difference is that,
from the weak-coupling expansion (or ``perturbative'' 
expansion) on the lattice, one obtains that the ghost-gluon vertex is
given at tree level by \cite{Kawai:1980ja}
\be
\Gamma_{\mu}^{abc}(\widetilde{k},\widetilde{q})
          \; =\; i \, g_0 \, f^{abc} \hat{q}_{\mu}
\cos\left(\frac{\pi\,\widetilde{s}_{\mu}\, a}{L_{\mu}}\right)\,\, ,
\label{eq:lamb0}
\ee 
where $\widetilde{s} = \widetilde{k}+\widetilde{q}$.
Clearly, by taking the formal continuum limit $\,a \to 0\,$ of the
quantity $\, \Gamma_{\mu}^{abc}(\widetilde{k},\widetilde{q})
/ a \,$ one recovers
the continuum tree-level result (\ref{eq:gamma0lev}).
Then, in the case $\widetilde{k} = 0$, $\widetilde{s} = \widetilde{q}$,
eq.\ (\ref{eq:z1L}) is still valid
on the lattice if one sets
\be
\Xi(\widetilde{q}) \;=\; 
            \frac{-i}{\hat{q}^2} \,
              \sum_{\mu} \, \frac{\hat{q}_{\mu}}{
            \cos\left(\frac{\pi \,\widetilde{q}_{\mu}\, a}{L_{\mu}}\right)} \,
                         \Gamma_{\mu}(0,\widetilde{q})
\ee
with
\bea
\Gamma_{\mu}(0,\widetilde{q}) & = & \frac{1}{g_0 \, N_c  \,(N_c^2 \, - \, 1)} \,
        \sum_{a,b,c}\, f^{abc}\,\Gamma^{abc}_{\!\mu}(0,\widetilde{q}) \\[2mm]
\Gamma^{abc}_{\!\mu}(0,\widetilde{q}) & = &
     \frac{G^{abc}_{\!\mu}(0,\widetilde{q})}{D(0) \, G^2(\widetilde{q})}
\eea
and
\be
\hat{q} \;=\;
  \left( \, \sum_{\mu = 1}^4\, \hat{q}_{\mu}^2 \,\right)^{1/2} \,\, ,
\ee
where $\hat{q}_{\mu}$ is defined in terms of
$\widetilde{q}_{\mu}$ in eq.\ (\ref{eq:platt}).
Thus, as in the continuum, we can write
\be
\Xi(\widetilde{q}) \;=\; \frac{1}{g_0 \, N_c  \,(N_c^2 \, - \, 1)} \;
     \frac{1}{\hat{q}^2}\;
              \sum_{\mu}\; \frac{\hat{q}_{\mu}}{
              \cos\left(\frac{\pi \,\widetilde{q}_{\mu}\, a}{L_{\mu}}\right)} \,
                       \sum_{a,b,c}\, f^{abc}\;   
      \mbox{Im}\; \Gamma^{abc}_{\!\mu}(0,\widetilde{q}) \,\, .
\ee

In order to check for possible effects due to the breaking of
rotational symmetry, in our simulations we use two types of
lattice momenta, i.e.\ $\,\widetilde{q}_1 = \widetilde{q}_2
= \widetilde{q}_3 = 0, \,
\widetilde{q}_4 = \widetilde{q}$ and $\widetilde{q}_1 = \widetilde{q}_2 =
\widetilde{q}_3 = \widetilde{q}_4 = \widetilde{q}$.
In the following we will indicate the first type of momenta as
{\em asymmetric} and the second one as {\em symmetric}.
Clearly, using the relations $L_1 = L_2 = L_3 = L_4 = L$, we have
\be
\hat{q} \;=\; \hat{q}_4 \;=\;
     2 \, \sin{\left(\frac{\pi\, \widetilde{q} \, a}{L}\right)}
\ee
in the asymmetric case and
\bea
\hat{q}_{\mu} &=& 2 \, \sin{\left(\frac{\pi\, \widetilde{q} \, a}{L}\right)}
       \\[2mm]
\hat{q} &=&
     4 \, \sin{\left(\frac{\pi\, \widetilde{q} \, a}{L}\right)}
\eea
in the symmetric one.
Then, the quantity $\Xi(\widetilde{q})$ above becomes
\be
\Xi(\widetilde{q}) \;=\; \frac{1}{g_0 \, N_c  \,(N_c^2 \, - \, 1)} \;
     \frac{1}{\hat{q}}\;
              \frac{1}{
              \cos\left(\frac{\pi \,\widetilde{q}\, a}{L}\right)} \,
                       \sum_{a,b,c}\, f^{abc}\;
      \mbox{Im}\; \Gamma^{abc}_{\!4}(0,\widetilde{q})
\ee
in the asymmetric case and
\be
\Xi(\widetilde{q}) \;=\; \frac{1}{g_0 \, N_c  \,(N_c^2 \, - \, 1)} \;
     \frac{1}{\hat{q}}\;
        \frac{1}{2 \, \cos\left(\frac{\pi \,\widetilde{q}\, a}{L}\right)} \,
              \sum_{\mu}\; \sum_{a,b,c}\, f^{abc}\;
      \mbox{Im}\; \Gamma^{abc}_{\!\mu}(0,\widetilde{q})
\ee
in the symmetric one.
Using the trigonometric relation $\sin(2 \,\theta) \,=\, 2\,\sin(\theta) \,
\cos(\theta)$ these formulae can be rewritten as
\be
\Xi(\widetilde{q}) \;=\; 
     \frac{1}{\sin\left(\frac{2 \, \pi \,\widetilde{q}\, a}{L}\right)}
                    \,\Sigma(\widetilde{q})
          \,\, ,
\label{eq:newxi}
\ee 
where
\be
 \Sigma(\widetilde{q}) \;=\; \frac{1}{g_0 \, N_c  \,(N_c^2 \, - \, 1)} \;
     \sum_{a,b,c}\, f^{abc}\;
      \mbox{Im}\; \Gamma^{abc}_{\!4}(0,\widetilde{q})
\label{eq:sigmaasym}
\ee
in the asymmetric case and
\be
 \Sigma(\widetilde{q}) \;=\; \frac{1}{g_0 \, N_c  \,(N_c^2 \, - \, 1)} \;
     \sum_{a,b,c}\, f^{abc}\; \mbox{Im}\;
      \frac{1}{4} \, \sum_{\mu} \; \Gamma^{abc}_{\!\mu}(0,\widetilde{q})
\label{eq:sigmasym}
\ee
in the symmetric one.

Finally, using eqs.\ (\ref{eq:eeA}), (\ref{eq:V}) and (\ref{eq:mm1})
we obtain
\be
G^{abc}_{\!\mu}(0,\widetilde{q}) \;=\; V\,\langle\,A_{\mu}^a(0)\,
                                     \left(M^{-1}\right)^{bc}(\widetilde{q})  \,
                                   \rangle \,\, ,
\ee
where
\be
A_{\mu}^a(0) \;=\; \frac{1}{V} \, \sum_{x}\; A_{\mu}^a(x)
\ee
and $ V $ is the lattice volume. Then, we can write
\be
\Gamma^{abc}_{\!\mu}(0,\widetilde{q}) \; = \;
     \frac{V\,\langle\,A_{\mu}^a(0)\,
                                     \left(M^{-1}\right)^{bc}(\widetilde{q})  \,
                                   \rangle}{D(0) \, G^2(\widetilde{q})} \,\, .
\label{eq:gammafin}
\ee
With our notation,
in the naive continuum limit $a \to 0$,
the lattice quantity $ 4 \, a^2 \, D(\widetilde{q}) / g_0^2$
[respectively $a^2 \, G(\widetilde{q})$] approaches the
(continuum) propagator
$ D(q^2) $ [respectively
$ G(q^2) $]. At the same time, we have that
$ 2 \, A_{\mu}^a(0) / ( a\, g_0) \to  A_{\mu}^a(0) $ 
and $a^2 \, \left(M^{-1}\right)^{bc}(\widetilde{q})
\to \left(M^{-1}\right)^{bc}(q)$.
This implies that the (normalized) lattice
ghost-gluon vertex $g_0\,\Gamma^{abc}_{\!\mu}(0,\widetilde{q}) / (2 \, a)$
approaches the quantity
$ \Gamma^{abc}_{\!\mu}(0,q) $ in the continuum limit.
Also, since $\hat{q} \to a \, q$ when $a \to 0$, it is
easy to verify that $\Z1^{-1}(\widetilde{q}) =
\Xi(\widetilde{q})$ is dimensionless.

\vskip 3mm

As said in the Introduction, in Ref.\ \cite{Bloch:2003sk}
the ghost-gluon renormalization function $\Z1(p^2)$ was
indirectly evaluated. To this end the authors considered
eq.\ (\ref{eq:alpha_run}), which can be written as
\be
Z_3(p^2, \beta) \, \widetilde{Z}_3^2(p^2, \beta) \;=\;
 \frac{\alpha_s(p^2)}{\alpha_0(\beta)} \; \Z1^2(p^2, \beta) \,\, ,
\ee
where the $\beta$ dependence is now indicated explicitly.
The renormalization functions $Z_3(p^2, \beta)$ and
$\widetilde{Z}_3^2(p^2, \beta)$ can be evaluated using the
matching technique described in detail in \cite[Sec.\ III]{Cucchieri:2003di}.
Then, one can show \cite{Bloch:2003sk}
that the left-hand side of the above equation rises linearly with
$-  \ln ( \sigma a^2) $, where $\sigma$ is the string tension.
At the same time, for large enough
$\beta$ values one finds that \cite{Bloch:2003sk}
\be
\frac{1}{\alpha_{0}(\beta)} \; \propto \;
- \, \ln ( \sigma a^2) \; + \; \mathrm{constant},
\ee
where the constant is $\beta$-independent.
It follows that $\widetilde{Z}^2_1(p^2, \beta)$ must be
finite in the continuum limit, if
the renormalized coupling $\alpha_s(p^2)$ is assumed finite.
Of course, this approach does not allow either an effective
determination of the vertex renormalization function or
a study of its $p$ dependence. These are the main goals of the
present work.


\section{Results}
\label{sec:results}

\FIGURE[t]{
\parbox{5.8in}{  
\hskip 0.1in
\epsfig{file=Sigma.eps,width=14cm}}
\caption{Results for $\Sigma(p)$
for the lattice volume $V = 16^4$ as a function of $p = \hat{p}/a$ in GeV,
considering asymmetric and symmetric momenta.
Error bars were obtained using the bootstrap method with 250 samples.
The dashed line represents the tree-level momentum dependence ($\sim
p$) of the vertex function in the continuum.
\label{fig:Sigma}}}

We have evaluated the reduced ghost-gluon vertex function
$\Gamma^{abc}_{\!\mu}(0,\widetilde{p}) $ at the
asymmetric point $(0; \widetilde{p}, -\widetilde{p})$
using eq.\ (\ref{eq:gammafin}), in the $SU(2)$ case.
(Let us recall that when $N_c = 2$ the structure functions
$f^{abc}$ are given by the completely anti-symmetric tensor
$\epsilon^{abc}$.)
Results for the quantity $\Sigma(\widetilde{p})$, defined in
eqs.\ (\ref{eq:sigmaasym}) and (\ref{eq:sigmasym}), and
for $\Z1^{-1}(\widetilde{p}) = \Xi(\widetilde{p}) $
[see eq.\ (\ref{eq:newxi})] as a function of
$p = \hat{p} / a$ (in physical units) are reported in
Figs.\ \ref{fig:Sigma} and \ref{fig:Z}. We consider for these
figures the lattice
volume $V = 16^4$, for both asymmetric and symmetric momenta.
Error bars were obtained using the bootstrap method with 250 samples.
(We checked that the results do not change when considering
500 samples.)
We find that the function $\Sigma(\widetilde{p})$ has the
same momentum dependence as the tree-level vertex
[i.e.\ $\sim \hat{p} \, \cos(\pi \, \widetilde{p} \, a / L) \sim
 \sin( 2 \, \pi \, \widetilde{p} \, a / L) $] and
that (consequently)
$\Z1(\widetilde{p})$ is approximately constant and equal to 1.

\FIGURE[t]{
\parbox{5.8in}{  
\hskip 0.03in
\epsfig{file=ASym_Z16.eps,width=7cm}
\hskip 5mm   
\epsfig{file=Sym_Z16.eps,width=7cm}}
\caption{Results for $\,\Z1^{-1}(p)\,$
for the lattice volume $V = 16^4$ as a function of $p = \hat{p}/a$ in GeV,
considering asymmetric (left) and symmetric (right) momenta.
In both cases we show data obtained considering
the two different gauge-fixing
methods (without and with smearing).
Error bars were obtained using the bootstrap method with 250 samples.
\label{fig:Z}}}

As explained in Section \ref{sec:data} above,
in order to check for Gribov-copy
effects we have considered two different gauge fixing methods
for each thermalized
configuration, the second of which employs the so-called
smeared gauge fixing.
Let us recall that Gribov-copy effects have been found
for the ghost propagator in Landau gauge,
at least in the small-momentum limit
\cite{Nakajima:2003my,Cucchieri:1997dx,Cucchieri:1997ug,Bakeev:2003rr}.
Recently, such effects were also found for the
gluon propagator in the infrared region \cite{Silva:2004bv}.
In Fig.\ \ref{fig:Z} are also shown the data for
$\Z1^{-1}(\widetilde{p})$ obtained
for the lattice volume $V = 16^4$ using the smeared gauge-fixing.
The $\Z1^{-1}(\widetilde{p})$ data for all
lattice volumes and $\beta$ values considered in this work are
reported in Tables \ref{table:z1asym} (asymmetric momenta) and
\ref{table:z1sym} (symmetric momenta) considering the
two different gauge-fixing methods (without and with smearing).
We can clearly see that $\Z1^{-1}(\widetilde{p})$ decreases
as the lattice volume increases for a fixed coupling $\beta$ and that
for $V = 16^4$
the dependence of $\Z1^{-1}(\widetilde{p})$ on the
coupling $\beta$, on the type of momenta and on Gribov-copy
effects is relatively small. This is also evident if we try to fit
the data for $\Z1^{-1}(\widetilde{p})$ to a constant (see Table
\ref{table:Z1fit}). Nevertheless, it is
clear that data obtained using the smeared gauge-fixing are
generally smaller than those obtained without smearing.

It is difficult to explain the small variations in the
results for $\Z1^{-1}(\widetilde{p})$ when considering
asymmetric and symmetric momenta or the smeared gauge fixing
compared to the standard gauge fixing.
Indeed, several quantities enter the definition of
$\Gamma^{abc}_{\!\mu}(0,\widetilde{p})$ [see eq.\ (\ref{eq:gammafin}]
and therefore the evaluation of $\Z1^{-1}(\widetilde{p})$.
A more detailed study
(see Tables \ref{tab:D0}, \ref{tab:Gsmall} and
\ref{tab:vertexasym})
of the gluon propagator at
zero momentum $D(0)$, the ghost propagator
$G(\widetilde{p})$ for
$\widetilde{p}_1 = \widetilde{p}_2 = \widetilde{p}_3 = 0,
\widetilde{p}_4 = N/2$ and the quantity
$ \sum_{a,b,c}\, f^{abc}\; \mbox{Im}\; 
 \langle\,A_{\mu}^a(0)\, \left(M^{-1}\right)^{bc}(\widetilde{p}) \,
                                   \rangle$
as a function of $p = \hat{p} / a$ does not
clarify the situation. In particular, the Gribov-copy effects,
if present, are always
small and within error bars. This is not surprising,
since we are considering relatively small lattice volumes
in the scaling region.
We can conclude that the small systematic variations in the
results of $\Z1(\widetilde{p})$ are probably related to
correlations among data at different momenta, since they have
been obtained using the same set of configurations.

Finally, we can use our data to evaluate the running
coupling constant $\alpha_s(p^2)$ using eq.\ (\ref{eq:alpha_run}).
For $\Z1(p^2)$ we take the fit to a constant
reported in Table \ref{table:Z1fit}.
The results are shown in Fig.\ \ref{fig:alpha}. We can see
that data obtained at different $\beta$ values do not lie
on a single curve. This is probably related to the fact that
our data are not at infinite volume. Indeed, this is a
well-known effect for the gluon field when evaluated at
zero momentum \cite{Cucchieri:1997fy,Cucchieri:1999sz,
Cucchieri:2000gu,Cucchieri:2000kw,Cucchieri:2001tw}.

\FIGURE[t]{
\parbox{5.8in}{  
\hskip 0.02in
\epsfig{file=ASym_alpha_new.eps,width=7cm}
\hskip 5mm   
\epsfig{file=Sym_alpha_new.eps,width=7.1cm}}
\caption{Results for the running coupling $\alpha_s(p)$
as a function $p = \hat{p}/a$ in GeV,
considering asymmetric (left) and symmetric (right) momenta.
Here we use the standard gauge-fixing method (without smearing).
Error bars were obtained using the bootstrap method with 250 samples.
\label{fig:alpha}}}


\section{Conclusions}
\label{sec:conclusions}

We have presented the first numerical study of
the reduced ghost-gluon vertex function
$\Gamma^{abc}_{\!\mu}(0,p) $ and of the
renormalization function $\Z1(p^2)$ in minimal Landau gauge.
We have considered the $SU(2)$ case and
the asymmetric point $(0; p, -p)$.
We found that the vertex function has the same
momentum dependence of the tree-level vertex and that
$\Z1(p^2)$ is approximately constant and equal
to 1, at least for momenta $p \gtapprox 1$ GeV. This is 
a direct nonperturbative verification of the
well-known perturbative result
that $\Z1(p^2)$ is finite and constant
to all orders of perturbation theory
\cite{Taylor:ff,Piguet:1995er}. In particular, using the result obtained at the
largest value of $\beta$ considered here (i.e.\ $\beta = 2.4$)
we can write
$\Z1^{-1}(p^2) = 1.02^{+6}_{-7}$ (see Table \ref{table:Z1fit}),
where errors include Gribov-copy effects and discretization
errors related to the breaking of rotational invariance.

We are now extending this study, considering larger lattice
volumes (and therefore smaller momenta), the symmetric
point $k^2 = q^2 = s^2$, the $3d$ case and the $SU(3)$ gauge group. 
A significant improvement of our results for $\alpha_s(p^2)$ 
may be expected by considering the symmetric point.


\acknowledgments

This work was supported by
Funda\c{c}\~ao de Amparo \`a Pesquisa do Estado de S\~ao Paulo (FAPESP)
through grants: 00/05047--5 (AC, TM) and 03/00928--1 (AM).
Partial support from Conselho Nacional de Desenvolvimento
Cient\'{\i}fico e Tecnol\'ogico (CNPq) is also acknowledged
(AC, TM).


\TABULAR[t]{|c|c|c|c|c|c|c|c|}{
\hline 
$p$ (GeV)   & $\,$ 0.366  $\,$ &
            $\,$ 0.718  $\,$ &
            $\,$ 1.04   $\,$ &
            $\,$ 1.33   $\,$ &
            $\,$ 1.56   $\,$ &
            $\,$ 1.73   $\,$ &
            $\,$ 1.84   $\,$ \\
\hline
$N=4$       &   -    &    -   &    -   &  1.40(5)&    -    &     -   &    -   \\
$N=4$, sme  &   -    &    -   &    -   &  1.39(4)&    -    &     -   &    -   \\

\hline
$N=8$       &   -    & 1.08(5)&    -   & 1.11(4) &    -    & 1.13(5) &    -   \\
$N=8$, sme  &   -    & 1.10(5)&    -   & 1.12(4) &    -    & 1.14(4) &    -   \\

\hline
$N=16$      & 1.05(5)& 1.13(5)& 1.16(5)& 1.14(5)& 1.14(5) & 1.15(5) & 1.15(5)\\
$\,\,N=16$, sme$\,\,$ &
            0.98(5)& 1.06(5)& 1.07(5)& 1.06(5) & 1.05(5) & 1.05(5) & 1.05(5)\\
\hline
\hline 
$p$ (GeV)   & 0.466  & 0.913  & 1.33   & 1.69    & 1.98    & 2.20    & 2.34   \\

\hline
$N=4$       &   -    &    -   &    -   & 1.31(4) &    -    &     -   &    -   \\

$N=4$, sme  &   -    &    -   &    -   & 1.30(4) &    -    &     -   &    -   \\

\hline
$N=8$       &   -    & 1.30(5)&    -   & 1.22(5) &    -    & 1.21(5) &    -   \\

$N=8$, sme  &   -    & 1.24(6)&    -   & 1.18(5) &    -    & 1.18(5) &    -   \\

\hline
$N=16$      & 1.01(4)& 1.05(4)& 1.05(5)& 1.04(5) & 1.04(5) & 1.04(5) & 1.03(4)\\

$\,\,N=16$, sme$\,\,$ &
              1.06(5)& 1.09(5)& 1.08(5)& 1.07(5) & 1.07(5) & 1.07(5) & 1.06(5)\\
\hline
\hline 
$p$ (GeV)   & 0.644  & 1.26   & 1.83   & 2.34    & 2.75    & 3.05    & 3.24  \\
\hline
$N=4$       &   -    &    -   &    -   & 1.36(4) &    -    &     -   &    -   \\

$N=4$, sme  &   -    &    -   &    -   & 1.35(3) &    -    &     -   &    -   \\

\hline
$N=8$       &   -    & 1.16(5)&    -   & 1.06(5) &    -    & 1.06(4) &    -   \\

$N=8$, sme  &   -    & 1.15(5)&    -   & 1.07(5) &    -    & 1.06(4) &    -   \\

\hline
$L=16$      & 1.11(6)& 1.10(5)& 1.08(5)& 1.08(5)& 1.07(5) & 1.08(5) & 1.08(5) \\

$\,\,N=16$, sme$\,\,$ &
              1.08(6)& 1.06(5)& 1.04(5)& 1.05(5) & 1.05(5) & 1.05(5) & 1.05(5)\\

\hline
}{Results for
$\Z1^{-1}(\widetilde{p})$ as a function of $p = \hat{p}/a$ (in GeV)
for the three lattice sides $N = 4, 8, 16$ and the three $\beta$ values
considered (i.e.\ $2.2, 2.3$ and
$2.4$, respectively top, center and bottom part of the table) in the
case of asymmetric momenta, considering the two different gauge-fixing
methods (without and with smearing).
Error bars were obtained using the bootstrap method with 250 samples.
\label{table:z1asym}}

\clearpage

\TABULAR[t]{|c|c|c|c|c|c|c|c|}{
\hline 
$p$ (GeV)   & $\,$ 0.732  $\,$ &
            $\,$ 1.44   $\,$ &
            $\,$ 2.08   $\,$ &
            $\,$ 2.65   $\,$ &
            $\,$ 3.12   $\,$ &
            $\,$ 3.47   $\,$ &
            $\,$ 3.68   $\,$ \\
\hline
$N=4$       &   -    &    -   &    -   & 1.22(4) &    -    &     -   &    -   \\

$N=4$, sme  &   -    &    -   &    -   & 1.23(4) &    -    &     -   &    -   \\

\hline
$N=8$       &   -    & 1.07(5)&    -   & 1.01(5) &    -    & 0.99(5) &    -   \\

$N=8$, sme  &   -    & 1.03(5)&    -   & 0.98(5) &    -    & 0.97(5) &    -   \\

\hline
$N=16$      & 1.01(5)& 0.98(4)& 0.97(4)& 0.96(4) & 0.95(4) & 0.95(4) & 0.94(5)\\

$\,\,N=16$, sme$\,\,$ &
              0.99(4)& 0.97(4)& 0.96(4)& 0.95(4) & 0.95(4) & 0.94(4) & 0.94(5)\\

\hline
\hline 
$p$ (GeV)   & 0.931  & 1.83   & 2.65   & 3.37    & 3.97    & 4.41    & 4.68   \\

\hline
$N=4$       &   -    &    -   &    -   & 1.14(4) &    -    &     -   &    -   \\

$N=4$, sme  &   -    &    -   &    -   & 1.12(4) &    -    &     -   &    -   \\

\hline
$N=8$       &   -    & 1.13(6)&    -   & 1.05(5) &    -    & 1.03(5) &    -   \\

$N=8$, sme  &   -    & 1.03(5)&    -   & 0.97(5) &    -    & 0.95(5) &    -   \\

\hline
$N=16$      & 1.04(6)& 1.00(4)& 0.99(5)& 0.99(6) & 0.98(5) & 0.98(5) & 0.99(5)\\

$\,\,N=16$, sme$\,\,$ &
              1.01(5)& 0.97(5)& 0.96(5)& 0.95(4) & 0.94(4) & 0.94(5) & 0.94(5)\\

\hline
\hline 
$p$ (GeV)   & 1.29   & 2.53   & 3.67   & 4.67    & 5.49    &  6.10   & 6.48  \\
\hline
$N=4$       &   -    &    -   &    -   & 1.21(4) &    -    &     -   &    -   \\

$N=4$, sme  &   -    &    -   &    -   & 1.16(4) &    -    &     -   &    -   \\

\hline
$N=8$       &   -    & 1.17(7)&    -   & 1.05(5) &    -    & 1.03(5) &    -   \\

$N=8$, sme  &   -    & 1.17(7)&    -   & 1.04(4) &    -    & 1.02(5) &    -   \\

\hline
$N=16$      & 1.04(5)& 1.00(5)& 1.00(5)& 1.00(5) & 1.00(5) & 0.99(5) & 1.00(5)\\

$\,\,N=16$, sme$\,\,$ &
              0.97(5)& 0.94(5)& 0.95(5)& 0.94(6) & 0.94(5) & 0.94(5) & 0.94(5)\\

\hline
}{Results for
$\Z1^{-1}(\widetilde{p})$ as a function of $p = \hat{p}/a$ (in GeV)
for the three lattice sides $N = 4, 8, 16$ and the three $\beta$ values
considered (i.e.\ $2.2, 2.3$ and
$2.4$, respectively top, center and bottom part of the table) in the
case of symmetric momenta, considering the two different gauge-fixing
methods (without and with smearing).
Error bars were obtained using the bootstrap method with 250 samples.
\label{table:z1sym}}

\TABULAR[t]{|c|c|c|}{
\hline
      & asymmetric momenta & symmetric momenta   \\
\hline
$\,\,\beta =$ 2.2       
      $\,\,$ & $\,\,$ 1.148(4) $\,\,$ & $\,\,$ 0.960(6) $ \,\,$ \\
$\,\,\beta =$ 2.2, sme  
       $\,\,$ & $\,\,$ 1.056(4) $\,\,$ & $\,\,$ 0.952(5) $ \,\,$ \\
\hline
$\,\,\beta =$ 2.3       
       $\,\,$ & $\,\,$ 1.039(3) $\,\,$ & $\,\,$ 0.989(3) $ \,\,$ \\
$\,\,\beta =$ 2.3, sme  
       $\,\,$ & $\,\,$ 1.070(3) $\,\,$ & $\,\,$ 0.949(5) $ \,\,$ \\
\hline
$\,\,\beta =$ 2.4       
       $\,\,$ & $\,\,$ 1.082(4) $\,\,$ & $\,\,$ 1.004(6) $ \,\,$ \\
$\,\,\beta =$ 2.4, sme  
       $\,\,$ & $\,\,$ 1.050(3) $\,\,$ & $\,\,$ 0.946(4) $ \,\,$ \\
\hline
}{Constant fits of the
$\Z1^{-1}(\widetilde{p})$ data reported in
Tables \ref{table:z1asym} and \ref{table:z1sym}
for the lattice volume $V = 16^4$ considering momenta
$p \geq 1$ GeV.
\label{table:Z1fit}}

\TABULAR[t]{|c|c|c|c|}{
\hline
      & V = \( 4^{4} \) & V = \( 8^{4} \)& V = \( 16^{4} \)   \\
\hline
$\,\,\beta =
   $ 2.2          $\,\,$ & $\,\,$ 5.33(6) $\,\,$ & $\,\,$ 11.6(3) $\,\,$ &
                    $\,\,$ 14.5(4) $\,\,$ \\
$\,\,\beta =
   $ 2.2, sme     $\,\,$ & $\,\,$ 5.38(6) $\,\,$ & $\,\,$ 11.8(2) $\,\,$ &
                    $\,\,$ 14.4(3) $\,\,$ \\
\hline
$\,\,\beta =
   $ 2.3          $\,\,$ & $\,\,$ 6.17(7) $\,\,$ & $\,\,$ 18.9(4) $\,\,$ &
                    $\,\,$ 31.7(7) $\,\,$ \\
$\,\,\beta =
   $ 2.3, sme     $\,\,$ & $\,\,$ 6.23(7) $\,\,$ & $\,\,$ 19.3(4) $\,\,$ &
                    $\,\,$ 32.4(7) $\,\,$ \\
\hline
$\,\,\beta =
   $ 2.4          $\,\,$ & $\,\,$ 6.87(8) $\,\,$ & $\,\,$ 23.5(5) $\,\,$ &
                    $\,\,$ 53(1) $\,\,$ \\
$\,\,\beta =
   $ 2.4, sme     $\,\,$ & $\,\,$ 6.94(8) $\,\,$ & $\,\,$ 23.8(5) $\,\,$ &
                    $\,\,$ 53(1) $\,\,$ \\
\hline
}{Results for the gluon propagator at zero momentum $D(0)$ (in
lattice units)
for the three lattice sides ($N = 4, 8, 16$) and the three $\beta$ values
(i.e.\ $2.2, 2.3$ and $2.4$) considering the two different gauge-fixing
methods (without and with smearing).
Error bars were obtained using the bootstrap method with 250 samples.
\label{tab:D0}}

\TABULAR[t]{|c|c|c|c|}{
\hline
      & V = \( 4^{4} \) & V = \( 8^{4} \)& V = \( 16^{4} \)   \\
\hline
$\,\,\beta =
   $ 2.2          $\,\,$ & $\,\,$ 0.353(1)$\,\,$ & $\,\,$ 0.3174(5)$\,\,$ &
                    $\,\,$ 0.3098(3)$\,\,$ \\
$\,\,\beta =
   $ 2.2, sme     $\,\,$ & $\,\,$ 0.352(1)$\,\,$ & $\,\,$ 0.3175(5)$\,\,$ &
                    $\,\,$ 0.3097(3)$\,\,$ \\
\hline
$\,\,\beta =
   $ 2.3          $\,\,$ & $\,\,$ 0.340(1)$\,\,$ & $\,\,$ 0.3075(4)$\,\,$ &
                    $\,\,$ 0.2998(2)$\,\,$ \\
$\,\,\beta =
   $ 2.3, sme     $\,\,$ & $\,\,$ 0.340(1)$\,\,$ & $\,\,$ 0.3075(4)$\,\,$ &
                    $\,\,$ 0.2997(2)$\,\,$ \\
\hline
$\,\,\beta =
   $ 2.4          $\,\,$ & $\,\,$ 0.334(1)$\,\,$ & $\,\,$ 0.3013(4)$\,\,$ &
                    $\,\,$ 0.2946(3)$\,\,$ \\
$\,\,\beta =
   $ 2.4, sme     $\,\,$ & $\,\,$ 0.333(1)$\,\,$ & $\,\,$ 0.3013(4)$\,\,$ &
                    $\,\,$ 0.2945(3)$\,\,$ \\
\hline
}{Results for the ghost propagator $G(p)$ (in lattice units),
evaluated at the asymmetric momentum with
components $\widetilde{p}_1 = \widetilde{p}_2 = \widetilde{p}_3 = 0$
and $\widetilde{p}_4 = N/2$,
for the three lattice sides ($N = 4, 8, 16$) and the three $\beta$ values
(i.e.\ $2.2, 2.3$ and $2.4$) considering the two different gauge-fixing
methods (without and with smearing).
Error bars were obtained using the bootstrap method with 250 samples.
\label{tab:Gsmall}}

\scriptsize

\TABULAR[t]{|c|c|c|c|c|c|c|c|}{
\hline 
$p$ (GeV)   & $\,$ 0.366  $\,$ &
            $\,$ 0.718  $\,$ &
            $\,$ 1.04   $\,$ &
            $\,$ 1.33   $\,$ &
            $\,$ 1.56   $\,$ &
            $\,$ 1.73   $\,$ &
            $\,$ 1.84   $\,$ \\
\hline
$N=16$      &   0.28(2)   & 0.019(1)    & 0.0041(2) & 0.00143(7) & 
                0.00062(3) & 0.00029(2) & 0.000122(6) \\
$\,\,N=16$, sme$\,\,$ &
                0.26(2)   & 0.0177(9)   & 0.0038(2) & 0.00131(7) & 
                0.00057(3) & 0.00027(1) & 0.000111(6) \\
\hline
\hline 
$p$ (GeV)   & 0.466  & 0.913  & 1.33   & 1.69    & 1.98    & 2.20    & 2.34   \\

\hline
$N=16$      &   0.46(3)   & 0.031(2)     & 0.0069(4) & 0.0025(1)  & 
                0.00113(6) & 0.00054(3)  & 0.00023(1)  \\
$\,\,N=16$, sme$\,\,$ &
                0.48(4)   & 0.032(2)     & 0.0073(4) & 0.0026(1)  & 
                0.00118(7) & 0.00057(3)  & 0.00024(1)  \\
\hline
\hline 
$p$ (GeV)   & 0.644  & 1.26   & 1.83   & 2.34    & 2.75    & 3.05    & 3.24  \\
\hline
$N=16$      &   0.74(7)   & 0.046(3)     & 0.0108(6) & 0.0040(2)  & 
                0.0018(1) & 0.00089(5)   & 0.00037(2)  \\
$\,\,N=16$, sme$\,\,$ &
                0.70(6)   & 0.045(3)     & 0.0106(6) & 0.0039(2)  & 
                0.0018(1) & 0.00087(5)   & 0.00037(2)  \\
\hline
}{Results for the quantity 
$ \sum_{a,b,c}\, f^{abc}\; \mbox{Im}\;
 \langle\,A_{\mu}^a(0)\, \left(M^{-1}\right)^{bc}(\widetilde{p}) \,
                                   \rangle$
as a function of $p = \hat{p}/a$ (in GeV)
for the lattice side $N = 16$ and the three $\beta$ values
considered (i.e.\ $2.2, 2.3$ and
$2.4$, respectively top, center and bottom part of the table) in the
case of asymmetric momenta, considering the two different gauge-fixing
methods (without and with smearing).
Error bars were obtained using the bootstrap method with 250 samples.
\label{tab:vertexasym}}

\end{document}